\documentclass[twocolumn,showpacs,preprintnumbers,amsmath,amssymb,superscriptaddress]{revtex4}
\usepackage{amssymb}
\usepackage{graphicx,color}
\usepackage{bm}

\begin{document}

\title{Mach cone induced by $\gamma$-triggered jets in high-energy heavy-ion collisions}

\author{Han-Lin Li}
\affiliation{Institute of Particle Physics, Hua-Zhong Normal
University, Wuhan 430079, China}
\author{Fu-Ming Liu}
\affiliation{Institute of Particle Physics, Hua-Zhong Normal
University, Wuhan 430079, China}
\author{Guo-Liang Ma}
\affiliation{Shanghai Institute of Applied Physics, Chinese
Academy of Sciences, P.O. Box 800-204, Shanghai 201800, China}
\author{Xin-Nian Wang}
\affiliation{Institute of Particle Physics, Hua-Zhong Normal
University, Wuhan 430079, China}
\affiliation{Nuclear Science Division MS 70R0319, Lawrence Berkeley National Laboratory, Berkeley, California 94720}
\author{Yan Zhu}
\affiliation{Institute of Particle Physics, Hua-Zhong Normal
University, Wuhan 430079, China}

\begin{abstract}

Medium excitation by jet shower propagation inside a quark-gluon plasma is studied within a linear
Boltzmann transport and a multiphase transport model.  Contrary to the naive expectation, it is the deflection of both the 
jet shower and the Mach-cone-like excitation in an expanding medium that is found to
gives rise to a double-peak azimuthal particle distribution with respect to the initial jet direction.
Such deflection is the strongest for hadron-triggered jets which are often produced close
to the surface of dense medium due to trigger-bias and travel against or tangential to the radial flow. 
Without such trigger bias, the effect of deflection on $\gamma$-jet showers and their 
medium excitation is weaker. Comparative study of hadron and $\gamma$-triggered
particle correlations can therefore reveal the dynamics of jet-induced medium excitation in high-energy 
heavy-ion collisions.

\end{abstract}


\maketitle

Strong jet quenching has been observed in experiments \cite{Adler:2003qi,Adler:2002tq,Adare:2009vd,Abelev:2009gu} 
at the Relativistic Heavy-ion Collider (RHIC) as a consequence of jet quenching or parton  energy loss in high-energy
heavy-ion  collisions \cite{Wang:1991xy}. The energy and momentum lost by a propagating parton will be 
carried by radiated gluons and recoiled medium partons which in turn will go through further interaction and 
eventually lead to collective
medium excitation such as supersonic waves or Mach cones \cite{CasalderreySolana:2004qm,Stoecker:2004qu}.
Indeed, Mach cones have been found in the solutions of
both hydrodynamic response \cite{Ruppert:2005uz,Chaudhuri:2005vc,Betz:2009su,Neufeld:2009ep,Qin:2009uh}
and linearized Einstein equations in string theory \cite{Chesler:2007sv,Gubser:2009sn} excited by a propagating jet.
Such collective excitation by a propagating jet is expected to be responsible for the observed
conic back-to-back (b2b) azimuthal dihadron \cite{Adams:2005ph,Adler:2005ee} and trihadron correlations \cite{:2008nda}
with a maximum opening angle of $\Delta\phi\approx 1$ (rad) relative to the backside
of a triggered high-$p_{T}$ hadron. However, hadron spectra from the 
freeze-out of the Mach cone in both hydrodynamics with realistic energy-momentum deposition by 
jets \cite{CasalderreySolana:2004qm,Chaudhuri:2005vc,Betz:2009su} and string calculations in 
the hydrodynamic regime \cite{Gubser:2009sn} fail to reproduce the observed conic azimuthal correlations. 
Such correlations on the other hand are observed in a multiphase transport (AMPT) Monte Carlo simulations \cite{Ma:2006fm} 
which could come from jet-induced wakes that are deflected
by a radially expanding medium \cite{Betz:2009su}.

Dihadrons  with a high-$p_{T}$ trigger are mostly dominated by
b2b jets that are produced close to the surface of the dense matter \cite{Zhang:2007ja} with the
awayside jets often traveling against or tangential to the radial flow. 
Deflection of these jet showers and associated Mach cones by the radial flow 
can lead to double-peaked hadron azimuthal correlations. On the other hand, high-$p_{T}$ $\gamma$'s 
are produced throughout  the volume of the dense matter \cite{Zhang:2009rn}. The effect of deflection should be reduced 
for $\gamma$-triggered jet showers after averaging over all possible production positions and propagation direction, leading 
to a weaker double-hump $\gamma$-hadron correlation as compared to dihadron correlation.

In this Letter, we will study medium excitation by a propagating jet shower 
using both a linear Boltzmann transport and AMPT model \cite{Zhang:1999bd}. We will illustrate
that while a Mach-cone-like excitation by a propagating jet in a uniform medium cannot give rise to
a conic distribution of the final partons, deflection of the jet shower and the
Mach-cone-like excitation in an expanding medium will result in a double-peaked azimuthal distribution as observed
in dihadron measurements \cite{Adams:2005ph,Adler:2005ee}. Because of the different geometric distributions 
and propagation direction of the initial produced jets, we will illustrate that $\gamma$-hadron and dihadron 
azimuthal correlation will be quantitatively different, depending on the value of jet-medium cross section. 
We therefore propose to use comparative study of $\gamma$-hadron and dihadron azimuthal correlations
to shed light on the dynamics of jet-induced Mach-cone-like excitation in high-energy  heavy-ion collisions. It can
also disentangle other mechanisms such as triangular flow \cite{Alver:2010gr} and hot spots \cite{Takahashi:2009na}
that contribute to dihadron but not $\gamma$-hadron azimuthal correlation, though they in general enhance jet-medium 
interaction and the resulting medium modification of $\gamma$-hadron and dihadron azimuthal correlations.

We first study the jet shower propagation and medium excitation through a linearized Monte
Carlo simulation of the Boltzmann transport equation
\begin{eqnarray}
 p _1  \cdot \partial f_1 (p_1 ) &=&  - \int {dp_2 } {dp_3 } {dp_4 }   (f_1 f_2  - f_3 f_4 ) 
 \left| {M_{12 \to 34} } \right|^2 \nonumber \\
 &\times& 
   (2\pi )^4 \delta ^4 (p_1  + p_2  - p_3  - p_4 ), 
   \label{eq:boltz}
\end{eqnarray}
including only elastic $1+2\rightarrow 3+4$ processes as given by the matrix elements $M_{12 \to 34}$, 
where $ dp_i \equiv d^3 p_i/[2E_i (2\pi )^3]$, $ f_{i}  =1/(e^{p_{i}\cdot u/T}  \pm 1)$ $(i=2,4)$ are thermal parton 
phase-space densities in a medium with local temperature $T$ and flow velocity $u=(1,\vec v)/\sqrt{1-v^{2}}$, 
$f_{i}=(2\pi)^{3}\delta^{3}(\vec p-\vec p_{i})\delta^{3}(\vec x-\vec x_{i}-\vec v_{i}t)$ $(i=1,3)$ are the jet shower parton
phase-space densities before and after scattering, and we neglect the quantum statistics in the final state 
of the scattering. We will consider quark propagation in a thermal medium and assume small 
angle approximation of the elastic scattering amplitude  $\left| {M_{12 \to 34} } \right|^{2} = C g^{4}(s^2  + u^2
)/(t + \mu^{2 })^2 $ with a screened gluon propagator, where $s$, $t$ and $u$ are Mandelstam variables, 
$C$=1 (9/4) is the color factor for quark-gluon (gluon-gluon) scattering and $\mu$ is the screening mass
which we consider here as a constant cut-off of small angle scattering. The corresponding elastic cross section 
is $d\sigma/dt=\left| {M_{12 \to 34} } \right|^{2}/16\pi s^{2}$.

We assume partons follow classical trajectories between scatterings and keep 
track of the propagation of both jet shower partons and the recoiled medium parton after each scattering.
Both the jet and recoiled medium parton are allowed to scatter again with other medium partons
as they further propagate. We determine the probability of jet-medium
scattering in each time step $\Delta t_{j}$ according to an exponential distribution
\begin{equation}
P_{i}=1-\exp[-\sum_{j}\Delta x_{j}\!\cdot\! u \,\sigma_{i}\rho(\vec x_{j},t_{j})],
\end{equation}
where $\sigma_{i}$ is the scattering cross section between jet and medium partons, the sum is over the time steps along
the classical trajectory since the last scattering and $\rho$ is the local medium parton density. For each scattering,
we keep record of both the leading jet parton ($p_{3}$) and the recoil medium parton ($p_{4}$). Thermal
partons with the initial momentum ($p_{2}$)  will be subtracted from final parton
phase-space density to account for the back-reaction in the Boltzmann transport equation.
The net parton phase-space density $\delta f(p)$ averaged over many events will be the medium 
excitation by the propagating jet. We call such Monte Carlo simulation a linearized Boltzmann jet transport 
since we neglect scatterings  between recoiled medium partons. This is a good approximation to 
the full Boltzmann jet transport as long as the medium excitation remains relatively small $\delta f(p) \ll f(p)$.

\begin{figure}
\centerline{\includegraphics[width=7.0cm]{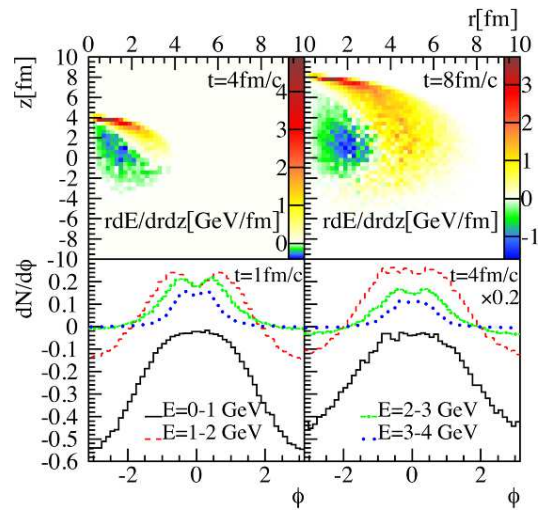}}
\caption{(Color online) Contour plot of energy density $rdE/drdz$ of medium excitation induced by
a quark with initial energy $E=20$ GeV that propagates in the $z$ direction in a uniform gluonic 
medium (upper panel) and the corresponding azimuthal parton distributions (lower panel).}
   \label{fig-uniform}
\end{figure}

Within such a linearized Boltzmann jet transport one can study not only  parton energy loss but also
the evolution of the medium excitation induced by the propagating jet. Shown in the upper panel of Fig.~\ref{fig-uniform} are
the contour plots of the energy density $rdE/drdz$ of the medium excitation induced by a quark
with initial energy $E=20$ GeV at different times after the start of the propagation along $z$ direction
in a uniform medium with temperature $T=300$ MeV.  We used $\alpha_{s}=0.5$ and $\mu=1.0$ GeV.
One can clearly see the formation of Mach-cone-like medium excitation at later times and 
a diffusion wake behind the jet which has negative energy density due to depletion of partons in this region by the
propagating jet, similar to results of the linearized hydrodynamic 
response \cite{CasalderreySolana:2004qm,Ruppert:2005uz,Neufeld:2009ep,Qin:2009uh} and 
string theory \cite{Chesler:2007sv,Gubser:2009sn} calculations.

We can also calculate parton spectra at different times,
corresponding to equal-time freeze-out in hydrodynamic studies. 
Shown in the lower panel of Fig.~\ref{fig-uniform} are the azimuthal distributions of partons from the medium
excitation at different times. At early times, the azimuthal
distribution is dominated by recoil thermal partons from the primary jet-medium scattering,
which is given by the collision kernel in the Boltzmann equation. It
has a double-hump feature in the azimuthal distribution for low energy partons with opening angles
determined by the value of temperature and screening mass $\mu$ in the scattering matrix element. As a parton propagates 
through the medium, such recoil thermal partons alway accompany it at the neck of the Mach-cone-like excitation.
This is very similar to the nonequilibrium part of jet-induced medium excitation in the string theory study \cite{Betz:2008wy}.

At later times, the recoil medium partons from the primary jet-medium interaction will scatter with other thermal
partons as they further propagate through the medium, causing diffusion of the wake front. As a consequence,
 low energy partons from the body of the Mach-cone-like excitation
overwhelm that from the neck and the final azimuthal distribution has only a broad single peak along the direction 
of the propagating jet.
 
The hot matter created in high-energy heavy-ion collisions has a finite initial transverse size as given
by that of two colliding nuclei. Because of the tremendous initial pressure, the hot matter will experience rapid
transverse expansion and develop strong radial flow. The nonuniformity over a finite transverse size and strong radial 
flow of the hot medium should influence the propagation of jet showers and their medium excitation. 

\begin{figure}[!ht]
\centerline{\includegraphics[width=7.0cm]{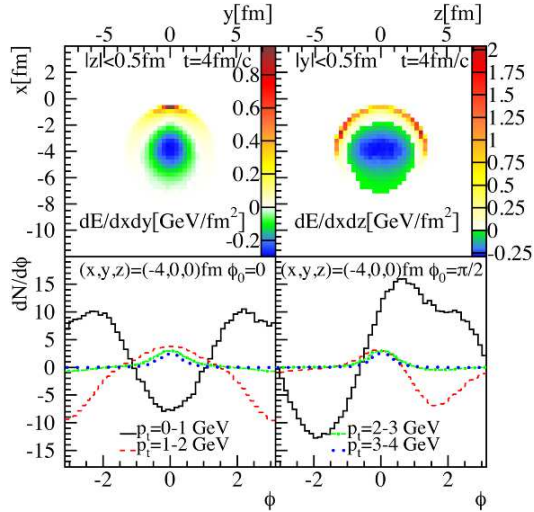}}
\caption{(Color online) (upper panel) Contour plot in the transverse ($x$-$y$) and beam ($x$-$z$) plane of energy density
excited by a quark jet shower with $E_{T}=20$ GeV and initial position at $(x,y,z)=(-4,0,0)$ fm that travels toward the center of
the expanding medium as given by ideal hydrodynamics \cite{Hirano:2005xf} for central $Au+Au$ collisions at the 
RHIC energy. The azimuthal distribution of medium and jet shower partons when the jet shower
travels against (lower left) and perpendicular (lower right) to the transverse flow.}
\label{fig-hydro}
\end{figure}

To take into account the dynamical evolution of the hot matter in our linearized Boltzmann jet
transport model, we will use the numerical results from a (3+1)D ideal hydrodynamical calculation \cite{Hirano:2005xf}
for local temperature and flow velocities, from which we generate thermal momentum $p_{2}$ and $p_{4}$ for 
each jet shower and medium parton interaction in Eq.~(\ref{eq:boltz}). We also use HIJING Monte Carlo model \cite{Wang:1991hta}
to provide the initial jet shower parton distribution which consists of multiple partons
from the final-state radiation in $\gamma$-jet events.  Shown in the upper panel of Fig.~\ref{fig-hydro} are the contour plots of the
energy density in both transverse ($x$-$y$) and the beam ($x$-$z$) plane excited by a $\gamma$-triggered jet 
shower with initial position $(x,y,z)=(-4,0,0)$ fm and energy $E^{\gamma}_{T}=20$ GeV in central $Au+Au$ collisions 
at $\sqrt{s}=200$ GeV/n. As compared to the case of a uniform medium, the  shape of the medium excitation is 
distorted considerably by the transverse and longitudinal flow of the expanding medium. 
The distortion depends on the direction of the jet propagation relative to the flow. 

The azimuthal distribution of partons from both the jet shower and jet-induced medium excitation is  
also distorted by the transverse flow and the nonuniformity of the dense medium. To illustrate the influence 
of transverse flow and the density gradient, we show in the lower panel of Fig.~\ref{fig-hydro} the
azimuthal distribution of jet shower and medium partons from a $\gamma$-triggered jet that is
produced at an initial position $(x,y,z)=(-4,0,0)$ fm
away from the center of the dense medium and propagating against (lower left) and perpendicular (lower right)  
to the radial flow, respectively. For a tangentially propagating jet shower (lower right), low $p_{T}$ partons from the jet shower and
Mach-cone-like excitation are clearly deflected by both the density gradience (we verify this by setting the 
transverse flow velocity to zero) and the radial flow, giving rise to the azimuthal distributions that peak 
at an angle away from the initial jet direction.  For jet showers 
that travel against the radial flow (lower left), the same deflection essentially splits the azimuthal
distribution of low $p_{T}$ partons to become a double-peaked one. Such 
deflection in an expanding system is also observed in an ideal hydrodynamical study of Mach cone 
propagation \cite{Betz:2009su} and it will give rise to both the diagonal (tangential jet showers)
and off-diagonal (split jet shower) part of the 3 particle correlation if analyzed as in the
experimental study \cite{:2008nda}. The magnitude of the conic correlations depends on the parton cross section
while the opening (deflection) angle also depends on the radial flow velocity.

To obtain the final jet-induced medium parton distribution in $\gamma$-jet events in heavy-ion collisions, one
should average over the propagation direction and the  initial production positions that are distributed in the
transverse plane according to the binary collisions with the nuclear geometry. The averaged
azimuthal distribution is found to have a broadened single peak in the jet direction for small values of the
parton cross section. The deflection is expected to give a double-peak azimuthal parton distribution 
for sufficiently large values of the parton cross section. However, within the
linearized Boltzmann jet transport model, larger parton cross sections will lead to larger amplitudes
of the medium excitation and eventually the linear Boltzmann transport breaks down which
neglects interaction among jet-excited medium partons. 

\begin{figure}[th]
\centerline{\includegraphics[width=7cm,height=7.0cm]{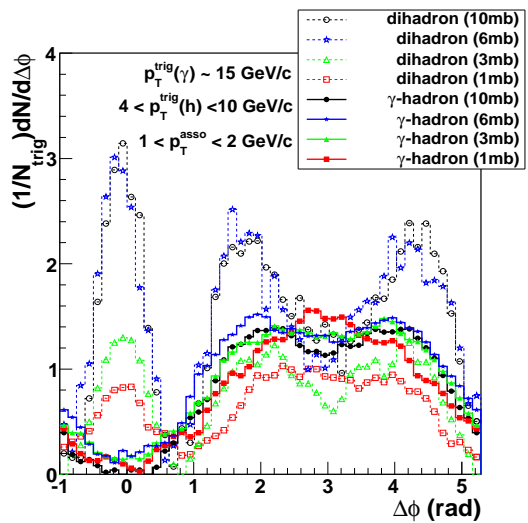}}
\caption{(Color online) Dihadron (open symbol) and $\gamma$-hadron (filled symbol) azimuthal  correlation 
from AMPT \cite{Zhang:1999bd} model calculation with different values of parton cross section.}
\label{fig-ampt1}
\end{figure}

\begin{figure}[th]
\centerline{\includegraphics[width=7cm,height=7.0cm]{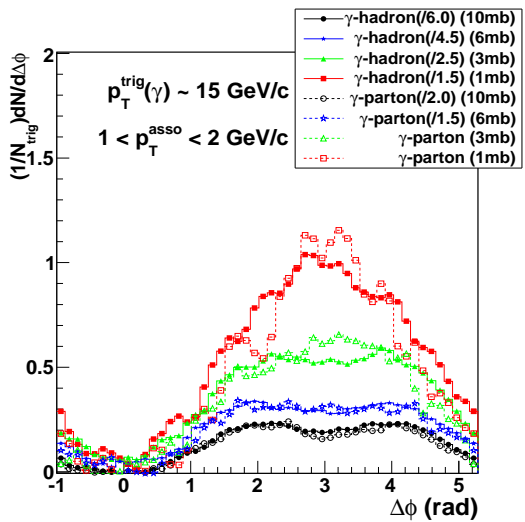}}
\caption{(Color online) $\gamma$-hadron  (filled symbol) and $\gamma$-parton (open symbol) azimuthal  correlation 
from AMPT \cite{Zhang:1999bd} model calculation with different values of parton cross section.  Results are
scaled with different factors for clear presentation}
\label{fig-ampt2}
\end{figure}

To go beyond the linear Boltzmann transport,
we use the AMPT Monte Carlo model to simulate hadron and $\gamma$-triggered jet events and study jet-induced
medium excitation. AMPT \cite{Zhang:1999bd} is essentially a full Boltzmann parton and hadron transport model
with only elastic parton collisions and initial conditions given by the HIJING model \cite{Wang:1991hta}. 
Shown in Fig.~\ref{fig-ampt1} are azimuthal distributions of hadrons (with $p_{T}^{asso}=1-2$ GeV/$c$)
associated with both a high-$p_{T}$ trigger $\gamma$ (closed symbols with solid lines) and 
hadron (open symbols with dashed lines)  in central $Au+Au$ collisions at the RHIC energy with 
different values of the parton cross section. The hadron correlations develop 
a double-peak feature in the opposite direction of the trigger due to the deflection of jet shower and 
Mach-cone-like excitation by the transverse flow and density gradient as one increases the value 
of parton cross section. The opening angle between two peaks increases with the value of the parton cross section.  
At very large values of the parton cross section, one should approach the hydrodynamic limit with
an opening $\Delta\phi\approx 1$ (rad) for the double-peak structure. Most importantly, one
can observe that  the amplitudes of the double-peaks in dihadron correlations are much bigger
than that of $\gamma$-hadron correlations for a given value of parton cross section, verifying our argument
that the deflection of jet shower and Mach-cone-like excitation by the radial flow has a stronger effect 
on the azimuthal dihadron correlation than $\gamma$-hadron correlation because of the difference in the trigger
bias on the initial jet production position and propagation direction. Therefore comparative study of dihadron 
and $\gamma$-hadron azimuthal correlations can shed light on the dynamics of jet propagation, medium
excitation, the strength of medium parton interaction and other effects such as hot spots \cite{Takahashi:2009na} 
and triangle flow \cite{Alver:2010gr} in high-energy heavy-ion collisions.

A hadronization mechanism such as parton coalescence employed in AMPT model can quantitatively 
influence the final hadron correlation \cite{Greco:2009ku} as compared to the parton correlation 
before hadronization. As shown in Fig.~\ref{fig-ampt2} such an effect, as manifested in the differences between 
$\gamma$-hadron and $\gamma$-parton correlation before hadronization, is rather small in AMPT.
We have not included inelastic parton interaction in this study but expect the results remain qualitatively the same. 
However, for quantitative theoretical predictions and comparison to experimental data, one should include inelastic 
parton interaction and hadronic interaction in future studies. In addition, the finite formation time 
for the jet parton shower, which is neglected in this study, will delay the jet shower and medium parton
interaction and quantitatively influence the final jet-induced $\gamma$-hadron and dihadron correlations.


We thank T. Hirano for providing the numerical results
of hydrodynamic calculations. This work is supported
by the NSFC under Projects No. 10610285, No. 10635020, No. 10705044, No. 10825523, No. 10975059 and 
by the U.S. DOE under Contract No. DE-AC02-05CH11231 and within the framework of the JET Collaboration.

\end{document}